\documentclass[aps,prd,twocolumn,floatfix,superscriptaddress]{revtex4}
\usepackage{amsmath}
\usepackage[dvips]{graphicx}
\newlength{\colw}
\newcommand{\Dslash}{\slash\hspace{-2.5mm}D}
\newcommand{\err}[2]{\raisebox{-0.4ex}
{$\stackrel{\scriptstyle +#1}{\scriptstyle -#2}$}}
\newcommand{\tmin}{t_{\text{min}}}
\newcommand{\tmax}{t_{\text{max}}}

\begin{document}

\title{Dynamical QCD simulations on anisotropic lattices}

\author{Richie Morrin}
\author{Alan \'O Cais}
\author{Mike Peardon}
\author{Sin\'ead M. Ryan}
\author{Jon-Ivar Skullerud}
\affiliation{School of Mathematics, Trinity College, Dublin~2, Ireland}
\collaboration{TrinLat Collaboration}

\date{\today}
\preprint{TrinLat-06/02}

\begin{abstract}
The simulation of QCD on dynamical ($N_f=2$) anisotropic lattices is
described. A method for nonperturbative renormalisation of the
parameters in the anisotropic gauge and quark actions is presented.
The precision with which this tuning can be carried out is determined
in dynamical simulations on $8^3 \times 48$ and $8^3 \times 80$ 
lattices. 
\end{abstract}

\maketitle

\section{Introduction}
\label{sec:intro}
The advantages of simulations with anisotropic lattices are well understood and the method has been 
used for precision determinations of an extensive range of quantities in the quenched approximation to 
QCD \cite{Karsch:1982ve,Manke:1998qc,Morningstar:1999rf,Asakawa:2003re,Hashimoto:2003fs,Ishii:2005vc}. 
In general a 3+1 anisotropy is employed where the 
lattice spacing in the temporal direction, $a_t$, is made fine whilst keeping the spatial lattice 
spacing $a_s$ relatively coarse. The advantages of this approach are two-fold. The improved resolution 
in the temporal direction means that states whose signal to noise ratio falls rapidly can be more 
reliably determined. The high computational cost of this improvement is offset by savings in 
the coarse spatial directions. 

The isotropic lattice (whose spacing in the four space-time directions is $a_x =
a_y = a_z = a_t \equiv a$) regulates QCD in a way that breaks the continuous
Euclidean symmetry down to the finite group of rotations of the hypercube.
Luckily the relevant operators that transform trivially under these two groups are the
same and so there is no renormalisation of the speed of light on the isotropic lattice.
Once an explicitly anisotropic lattice action is introduced with $a_x = a_y =
a_z \equiv a_s $ and $a_t \ne a_s$, the rotational symmetry of the theory is the cubic point group. For the
gluons, there are now two distinct operators not related by rotations at
dimension four: $\left\{\mbox{Tr }E^2, \mbox{Tr }B^2\right\}$; while for the quarks the set of dimension four operators
$\left\{ \bar\psi \;\Dslash\psi, \; m \bar\psi\psi \right\}$ grows to a set with three members:
$\left\{ \bar\psi \gamma_i D_i \psi, \; \bar\psi \gamma_0 D_0 \psi, \; m \bar\psi\psi \right\}$.
As a result, two new parameters appear in the action, and for the continuum
limit to represent QCD these parameters must be determined 
such that a physical probe of the vacuum at scales well below the cut-off appears to have full Euclidean
symmetry. 
The nonperturbative determination of these extra action parameters is the subject of the present paper.

In quenched QCD the anisotropy in the gauge sector, $\xi_g$, and the quark
sector, $\xi_q$, can be tuned independently and {\it post hoc} using two separate
criteria. The precision and mass-dependence of the determination of $\xi_q$ was
investigated for the action we use in Ref.~\cite{Foley:2004jf}. It was found that this parameter
could be determined at the percent level from the 
energy-momentum dispersion relation. The mass dependence was found
to be mild for quark masses in the range $m_s\leq m_q\leq m_c$ when the tuning
was carried out at the strange quark mass, $m_s$.
In Refs.~\cite{Klassen:1998ua,Alford:2000an}, a determination of the gluonic parameter was made to similar
precision. 

We would like to use anisotropic lattices in simulations with $N_f=2$
for realistic phenomenologically-relevant calculations. In dynamical
QCD the tuning procedure becomes more complicated because of the
interplay between the quark and gluon sectors and the parameters must
be simultaneously determined. There are several issues to
resolve. Firstly, can this simultaneous tuning be accomplished;
secondly, to what precision is the renormalised anisotropy determined;
and thirdly, what is the mass-dependence of the renormalised
anisotropy.  Here we will focus on the first two issues, and leave the
question of the mass dependence to a later study.

The paper is organised as follows. Section~\ref{sec:actions} gives the details of the gauge and quark actions 
used in this investigation. Section~\ref{sec:method} describes the tuning methodology and is followed in 
Section~\ref{sec:results} by the results for the values of the
tuned bare (input) parameters $\xi^0_g$ and $\xi^0_q$. 
Section~\ref{sec:conclusions} contains our conclusions and future plans. 

\section{The action and parameters}
\label{sec:actions}
We begin with a brief description of the anisotropic action used in this study. The details of the tuning 
procedure described in Section~\ref{sec:method} do not depend on the specific action used. Further description 
of the action can be found in \cite{Foley:2004jf} where the tuning for the same action in the quenched 
approximation was discussed. 

The gauge action is a two-plaquette Symanzik-improved
action~\cite{Morningstar:1999dh} previously developed for
high-precision glueball studies and given by
\begin{equation}
\begin{split}
  S_{G} =& \frac{\beta}{\xi^0_g} \left\{ \frac{5(1+\omega)}{3u^4_s}\Omega_s - 
             \frac{5\omega}{3u^8_s}\Omega^{(2t)}_s - \frac{1}{12u^6_s}\Omega^{(R)}_s \right\} \\
         & +\beta\xi^0_g \left\{ \frac{4}{3u^2_s u^2_t} \Omega_{t} - \frac{1}{12u^4_s u^2_t} 
             \Omega^{(R)}_t \right \} ,
\end{split}
\end{equation}
where $\Omega_s$ and $\Omega_t$ are spatial and temporal plaquettes. 
$\Omega_s^R$ and $\Omega_t^R$ are $2 \times 1$ rectangles in the $(i,j)$ and $(i,t)$ planes respectively. 
$\Omega_s^{2t}$ is constructed from two spatial plaquettes separated
by a single temporal link.  $u_s$ and $u_t$ are the mean spatial
and temporal gauge link values respectively.
The action has leading discretisation errors of ${\cal O}(a^3_s,a_t, \alpha_s a_s)$. 

For fermions an action specifically designed for large anisotropies is used. The usual Wilson term removes 
doublers in the temporal direction whereas spatial doublers are removed by the addition of a Hamber--Wu term. 
The action has been described in detail in Ref.~\cite{Foley:2004jf} and has leading classical discretisation 
errors of ${\cal O}(a_tm_q)$. In terms of continuum operators, it can be written
\begin{equation}
S = \bar{\psi}(\Dslash + m_0)\psi -\frac{ra_t}{2}\bar{\psi}\left ( D_0^2\right )
               \psi   +sa^3_s\bar{\psi}\sum_iD_i^4\psi ,
\end{equation}
which highlights the different treatment of temporal and spatial
directions. $r$ is the 
usual Wilson coefficient which is applied in the temporal direction only in this action and is set to unity. 
The analagous parameter in the spatial directions is $s$, 
which parameterises a term that is irrelevant in the continuum limit. A precise 
tuning of this parameter is not necessary: in practice we choose
$s=1/8$, so that the energy of a propagating quark 
at tree level increases monotonically across the Brillouin zone. 
Stout-link smearing \cite{Morningstar:2003gk} was used for the gauge fields in
the fermion matrix. Two stoutening iterations were used, with a parameter
$\rho=0.22$. This was fixed for all simulations, and chosen to approximately
maximise the expectation value of the spatial plaquette on the stout links. 

%

This study was carried out on $8^3\times 48$ and $8^3\times80$
anisotropic lattices with a spatial lattice spacing $a_s\approx0.2$fm
and a target anisotropy $\xi=6$. The bare sea quark mass was set to
$a_tm_q=-0.057$ in all runs. 
A set of gauge configurations, distributed across ten independent
Markov chains, was generated for each set of input 
parameters ($\xi^0_g$,$\xi^0_q$). 
Valence quark propagators were generated with the same mass as the sea
quarks.

To determine the statistical uncertainties, 1000 bootstrapped sets of
configurations were taken and analysis was done on these bootstrapped
sets.  Both point and all-to-all propagators were used.  Some
preliminary results using point propagators on $8^3\times48$ lattices
were presented in Ref.~\cite{Morrin:2005tc}.

\section{Methodology}
\label{sec:method}
The bare parameters, $\xi^0_g$ and $\xi^0_q$, are renormalised by demanding that
physical probes exhibit euclidean symmetry. In
principle, any physical quantity can be used; however, it should be easily
determined to high precision. In this study we have used the sideways potential
and the pion energy-momentum dispersion relation for the gauge and fermion
sectors respectively. 

The gauge anisotropy $\xi_g$ is determined from the interquark
potential~\cite{Klassen:1998ua,Alford:2000an}. The static source propagation is chosen to be 
along a coarse direction allowing the sources to be separated along both course and 
fine axes. The potential is determined  at the same physical distance for these two cases.
The input anisotropy is constrained so that the two calculations yield the same 
value of the potential, $V_s(x) = V_t(t/\xi)$ for a target anisotropy $\xi$.  For a
given input anisotropy $\xi^0_g$ and target anisotropy $\xi$ we can
determine the mismatch parameter $c_g=V_s(x)/V_t(t/\xi)$.  If $x$ is
in the r\'egime where the potential is nearly linear, the mismatch
parameter is approximately related to the actual gauge 
anisotropy, $c_g\approx\xi_g/\xi$. 

The quark anisotropy can be determined from the pseudoscalar
dispersion relation. The anisotropy is inversely 
proportional to the square root of the slope of the dispersion relation and demanding a relativistic 
energy-momentum relation imposes a renormalisation condition on the
bare parameter $\xi^0_q$. The ground 
state energy $E_0$ was determined for a range of momenta, $n^2 \in\{0,1,2,3,4,5,6\}$, where 
$p_n = \frac{2 \pi n}{L a_s}$ and we average over equivalent momentum values. 
The two-point correlator data were modelled with single exponentials and a 
$\chi^2$-minimisation was used to determine the best-fit ground state. 
These values were used to 
generate an energy-momentum dispersion relation. 

In the quenched approximation this procedure is relatively easy since $\xi^0_g$ and $\xi^0_q$ can be determined 
independently. For dynamical simulations it is no longer possible to simply fix $\xi^0_g$ and then tune 
$\xi^0_q$ to a consistent value, since changing $\xi^0_q$ will affect the measurement of $\xi_g$. 
Explicitly, changing the value of $\xi^0_q$ necessitates  a regeneration of the background fields with the 
new value of $\xi^0_q$ which in turn will change the measured anisotropy $\xi_g$ of the background fields. 
The solution to this problem is a simultaneous two-dimensional tuning procedure~\cite{Peardon:2002sd}. 

A linear dependence on the parameters $\xi^0_g$ and $\xi^0_g$ was assumed for a small region. Three initial sets 
of configurations were generated and the renormalised anisotropy was determined.
Planes were defined for both output values of $\xi_g$ and $\xi_q$ i.e. values $\alpha,\beta,\gamma$ were 
found to satisfy $\xi_a = \alpha_a\xi^0_g + \beta_a\xi^0_q + \gamma_a$ for the
renormalised anisotropy $\xi_a, a=g,q$ measured for each 
input $(\xi^0_q,\xi^0_g)$. The intersection of these planes with the required (target) output value yields the
tuned point.  The statistical uncertainties were determined using
bootstrap resampling, with a common bootstrap ensemble used for all
measurements. When more than three simulation points were available a
plane was defined using a constrained-$\chi^2$ fit.

All observables were estimated using the Monte Carlo method. An ensemble of 250
gauge field configurations divided across 10 Markov chains was generated using the Hybrid Monte Carlo (HMC)
algorithm \cite{Duane:1987de}. 
Approximately 5000 CPU hours were needed in order to generate each set of 
configurations. 
The HMC algorithm can be used for these simulations without modification. One
observation serves to improve performance, however. HMC adds a set of
momentum variables conjugate to the gauge fields, but each conjugate
momentum can be added with a different gaussian variance without changing the
validity of the method. In isotropic simulations this is not a useful property,
and all momentum co-ordinates are chosen to have unit variance. For the
anisotropic lattice, the temporal and spatial gauge fields have different
interactions, and different momenta become useful. 
If the HMC hamiltonian is 
\begin{equation}
{\cal H} = \sum_x \left( \frac{1}{2\mu_t^2} \mbox{Tr } P_0^2(x) + 
  \sum_{i=1}^3 \frac{1}{2} \mbox{Tr } P_i^2(x) \right)
  +S[U] ,
\end{equation}
an extra tunable parameter, $\mu_t$ (the variance of the temporal link momenta), 
has been added to the algorithm which can be used to optimise acceptance by
the Metropolis test. 
This is equivalent to using two distinct integration step-sizes
for the spatial and temporal degrees of freedom. Some brief
numerical experiments suggest that a temporal leap-frog step-size smaller by a 
factor $\xi$ is close to optimal, and this is borne out by considerations of
free field theory. 
\section{Results}
\label{sec:results}
\begin{table}
\begin{tabular}{l|ccccc}
Run                 & 1        & 2         & 3        & 4         & 5         \\
\hline
$\beta$ & 1.51 & 1.528 & 1.514 & 1.544 & 1.522 \\
$\xi_q^0$ & 6.0 & 7.5 & 7.5 & 8.72 & 8.83 \\
$\xi_g^0$ & 8.0 & 7.0 & 8.0 & 6.65 & 7.44 
\end{tabular}
\caption{Input parameters for the five dynamical simulations performed
  in this tuning procedure.  The bare quark mass is $a_tm_q=-0.057$ for
  all runs.}
\label{tab:inputs}
\end{table}
The input anisotropy parameters used are given in
Table~\ref{tab:inputs}.  We started by choosing three points (Runs 1--3) in
the $(\xi^0_g,\xi^0_q)$ plane, and generated configurations at two further
points as a result of the tuning procedure.
The final tuning was performed on $8^3\times80$ lattices, using data
from runs 1, 4 and 5 as these spanned the largest area of the plane.

\subsection{Interquark Potential}
\label{sec:sideways}

The gluon anisotropy is determined from the static quark potential at
a selected distance $R$. In practice this is done by determining
the effective energy for the static quark--antiquark configuration at
separation $R$ at some time $T$.  It is then important to choose
values for $R$ and $T$ where the potential is well determined and the
value obtained for $c_g$ is stable with respect to small variations in
$R$ and $T$.  The same values for $R$ and $T$ must then be used for
all runs in order to have a consistent procedure.

Table~\ref{tab:gluonRT} shows $c_g$ for different $R$ and $T$, on the
$8^3\times80$ lattices.  We see that the values are generally quite
consistent for each run.  Looking more closely at the effective
potential for each $R$ as a function of $T$, we find that it has not
yet reached a plateau at $T=1$, while the value for $T=3$ is
consistent within errors with that for $T=2$.  We choose $(T,R)=(2,3)$
as our optimal parameters, since this yields reasonably small
statistical errors, while $R$ is large enough to be in the linear
r\'egime.
\begin{table}
\begin{tabular}{l|llllll}
      &\multicolumn{6}{c}{$c_g = V_s(x)/V_t(t/\xi )$ at different (T,R)} \\
Run   &(1,3)&(1,4)&(2,3)&(2,4)&(3,3)&(3,4)\\
\hline
1 &0.972(2)&0.959(3)&0.972(7)&0.965(13)&0.991(25)&1.13(8)\\
4 &0.951(2) &0.941(4)&0.945(8) &0.926(18)& 0.942(34)&0.89(9)\\
5 &0.994(2)&0.990(3)&0.991(7)&0.998(13)&0.965(25)&1.01(7)
\end{tabular}
\caption{The gluon anisotropy parameter $c_g$ for different separations, $R$ and 
times, $T$. The final results were determined from data at $T=2$ and $R=3$.}
\label{tab:gluonRT}
\end{table}

\subsection{Dispersion relations}
\label{sec:dispersion}

Pseudoscalar meson correlators were computed using traditional point
propagators as well as all-to-all propagators \cite{Foley:2005ac} with
time and colour dilution and no eigenvectors.

\begin{figure}[h]
        \centering
	\includegraphics*[width=\colw]{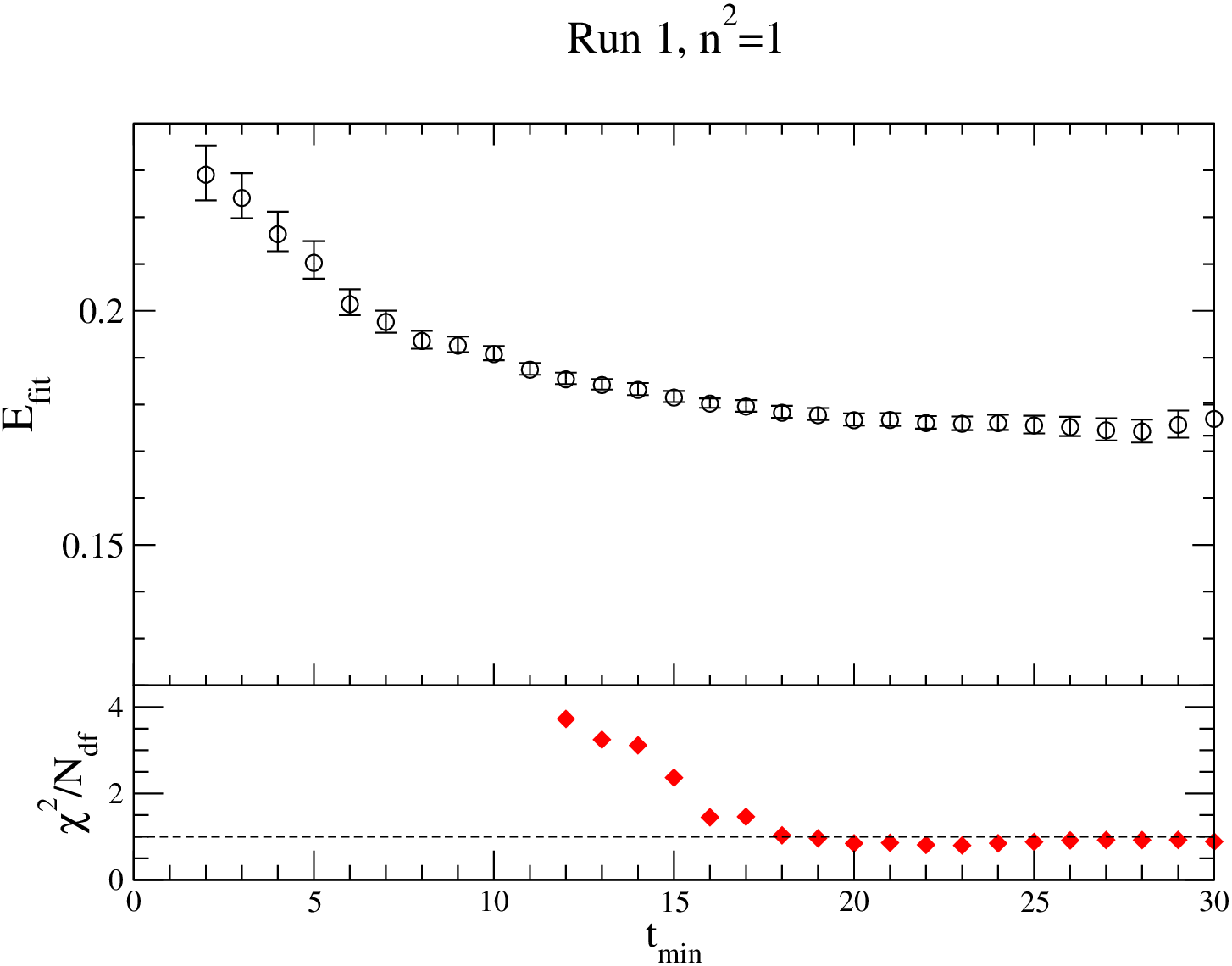}
	\caption{A typical $\tmin$ plot, showing the energy for momentum
          $n^2=1$ on run 1, $8^3\times 80$ lattices from fits to time ranges $\tmax=40$ for various
	  $\tmin$.  A stable ground state energy determination, with a
          good $\chi^2$, is achieved for
	  $22\leq\tmin\leq30$.}
	\label{fig:tmin}
\end{figure}

\begin{figure}[h]
  \centering
  \includegraphics*[width=\colw]{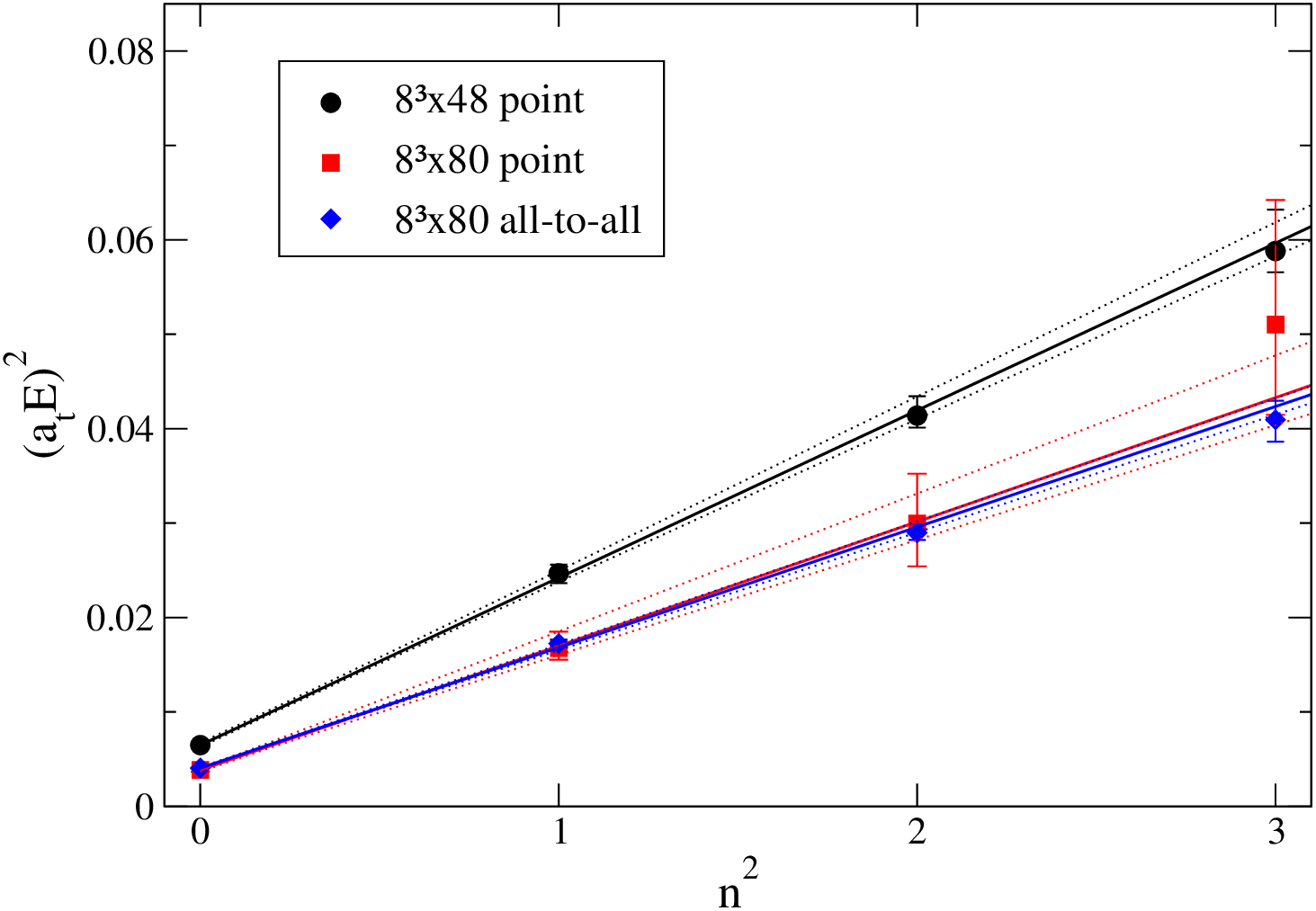}
  \caption{A comparison of the dispersion relations determined from an
    $8^3\times 48$ lattice and an $8^3\times 80$ lattice. The solid lines are the 
    best fits and the dotted lines are the 68\% confidence levels. The figure also
    shows a comparison of all-to-all propagators and point propagators on
    the same (longer) lattice. The plot shows that the ground state
    energies have not reached a plateau on the shorter lattice.  On the
    longer lattice the all-to-all and point data agree, while higher
    precision is achieved with all-to-all propagators.}
  \label{fig:longVSshort}
\end{figure}

\begin{figure}[ht]
  \centering
  \includegraphics*[width=\colw]{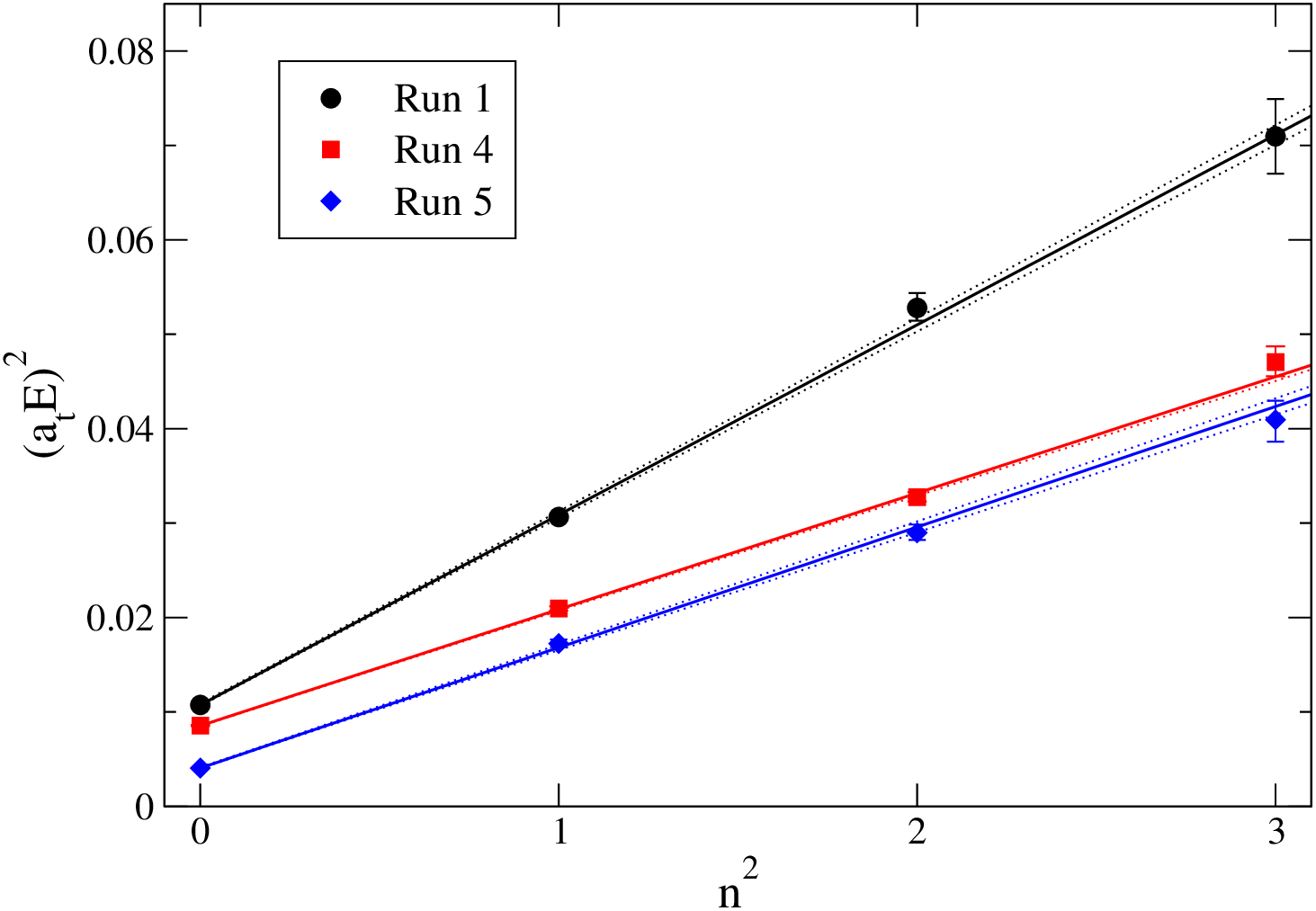}
  \caption{Dispersion relations from runs 1, 4 and 5 on $8^3\times80$
  lattices using all-to-all propagators. 
  The solid line is a fit to the four points and the dotted lines are the 
  68\% confidence levels. The quality of all three fits is very good with 
  $\chi^2/N_{d.f.} = 2.0/2, 1.9/2, 2.0/2$ for runs 1,4 and 5 respectively.}
  \label{fig:dispersion}
\end{figure}

To determine optimal fit
ranges for exponential fits to the correlator data, sliding window ($\tmin$) plots
were used: the correlation function was fitted in a range from $\tmin$ to 
$\tmax$ where $\tmax$ was fixed to the largest value compatible with a 
good fit, and $\tmin$ was varied.  An example of such a plot is given in Fig.~\ref{fig:tmin}.
The fit range was chosen so the fit would be stable with respect to 
small variations in $\tmin$.  The same fit ranges and smearing
parameters were chosen for all simulation points in order to obtain a
consistent determination of the dispersion relation.  The final fit ranges
are given in Table~\ref{tab:fitranges}.
\begin{table}
\begin{tabular}{l|lll}
$n^2$ & $\tmin$ & $\tmax$\\\hline
0   &  25   & 40 \\
1   &  24   & 40 \\
2   &  21   & 40 \\
3   &  19   & 40 \\
\end{tabular}
\caption{Fit ranges.}
\label{tab:fitranges}
\end{table}
In our initial analysis data from a $8^3\times 48$ lattice were used. However, 
a reliable extraction of the ground state energy proved difficult. In particular, 
it was observed that the energy either did not reach a plateau until near the end of the 
lattice or did not plateau at all. To resolve this problem the simulation was repeated
on a longer, $8^3\times 80$ lattice. An immediate improvement in the quality of the 
fits was observed. The ground state energy was determined from fits over at least 15 
timeslices and was stable with respect to changes in $\tmin$. The effect of the 
longer lattice is illustrated in Figure~\ref{fig:longVSshort}. This plot also compares 
simulations using point and all-to-all propagators. 
The all-to-all propagators lead to improved 
precision in the fitted energies. The central values are in agreement with the 
energies determined using point propagators but the statistical error is smaller. 
The final tuned parameters were determined using all-to-all propagators on the 
$8^3\times 80$ lattices. We find consistently good fits for all runs for the first four 
momenta considered ($n^2 = 0,1,2$ and 3). 
The renormalised quark anisotropy is therefore determined from fits to these momenta. 
Figure~\ref{fig:dispersion} shows the pseudoscalar dispersion relations for Runs 1, 4 and 5 which 
are used to determine the tuned point. 
\subsection{Plane fits}
\label{sec:planefits}
\begin{figure}
\includegraphics*[width=\colw]{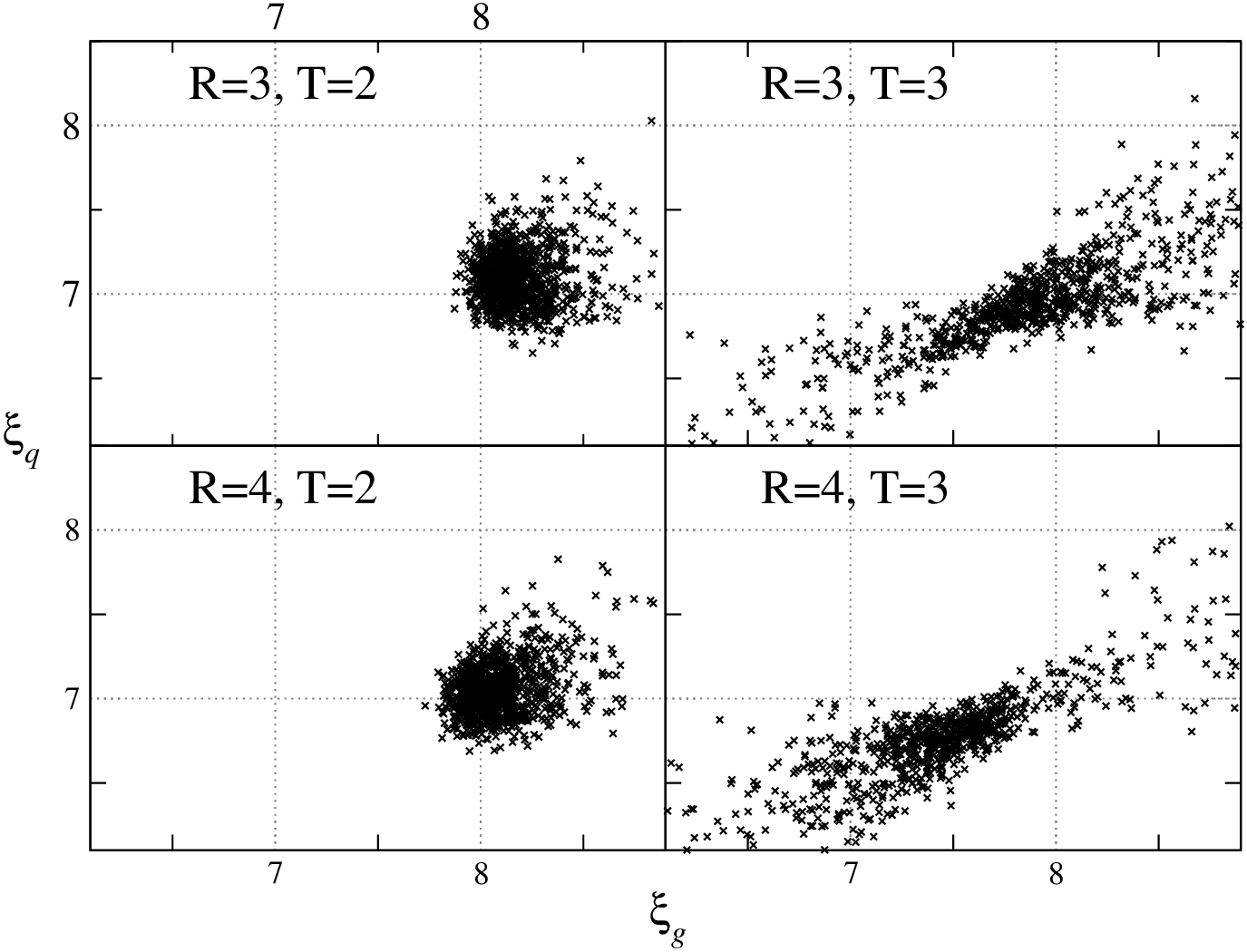}
\caption{Tuned values of input parameters $(\xi_g^0, \xi_q^0)$ determined from the plane fit 
procedure on the $8^3\times 80$ lattice. The plots show the results for different values of $R$ and
  $T$ used to determine the gluon anisotropy.  
Each point corresponds to one bootstrap sample.}
\label{fig:tuning-RT}
\end{figure}

Table~\ref{tab:out} shows the output anisotropies determined on the
$8^3\times48$ and $8^3\times80$ lattices for the five simulation points. 
\begin{table}[ht]
\begin{tabular}{c|cc|cc}
    &\multicolumn{2}{|c|}{$8^3\times 48$} & \multicolumn{2}{|c}{$8^3\times 80$} \\
\hline
Run & $c_g$ & $\xi_q$ & $c_g$ & $\xi_q$ \\
\hline
1   & 0.991(3) & 4.98(6) & 0.972(7) & 5.54(6) \\
2   & 0.986(3) & 6.27(4) &          &         \\
3   & 1.001(3) & 5.18(6) &          &          \\
4   & 0.985(5) & 6.47(5) & 0.945(8) & 7.08(5)  \\
5   & 0.995(3) & 5.80(5) & 0.991(7) & 6.95(8)  
\end{tabular}
\caption{Table of measured output anisotropies at each of the run points. The errors are statistical only.}
\label{tab:out}
\end{table}
As a check on the stability of our tuning procedure, we have repeated
the calculation using different values of $R$ and $T$ in the
determination of the gluon anisotropy.  The results are shown in
Fig.~\ref{fig:tuning-RT}. The plot shows that the anisotropies are 
insensitive to a change in $R$ but that increasing the value of $T$ from 
two to three leads to large statistical uncertainty, particularly in 
the gluon anistropy. For these reasons we choose $R=3$ and $T=2$ for 
our analysis.

\subsection{Simulation with tuned parameters}
Applying the plane fit procedure of Sec.~\ref{sec:planefits} to a
subset of configurations of Runs 1, 4 and 5 we obtained
preliminary, tuned parameters $\xi^0_g=8.06\err{7}{7},
\xi^0_q=7.52\err{21}{15}$.
250 configurations were generated with these parameters, and 
$c_g$ and $\xi_q$ determined using the same values for $R$, $T$ and fit ranges
as in Sections \ref{sec:sideways} and \ref{sec:dispersion}.  We find
$c_g=0.983(6), \xi_q=6.21(9)$. 
We see that both quark and gluon anisotropies are within 3\% of the
target value of 6.  Although the anisotropies are not equal within statistical
errors, we note that there are still
systematic uncertainties at the percent level, in particular for
$\xi_g$, as shown in Table~\ref{tab:gluonRT}.  For example, if we
choose $R=3, T=3$ we find $c_g=1.01(2)$.

\begin{figure}
\includegraphics*[width=\colw]{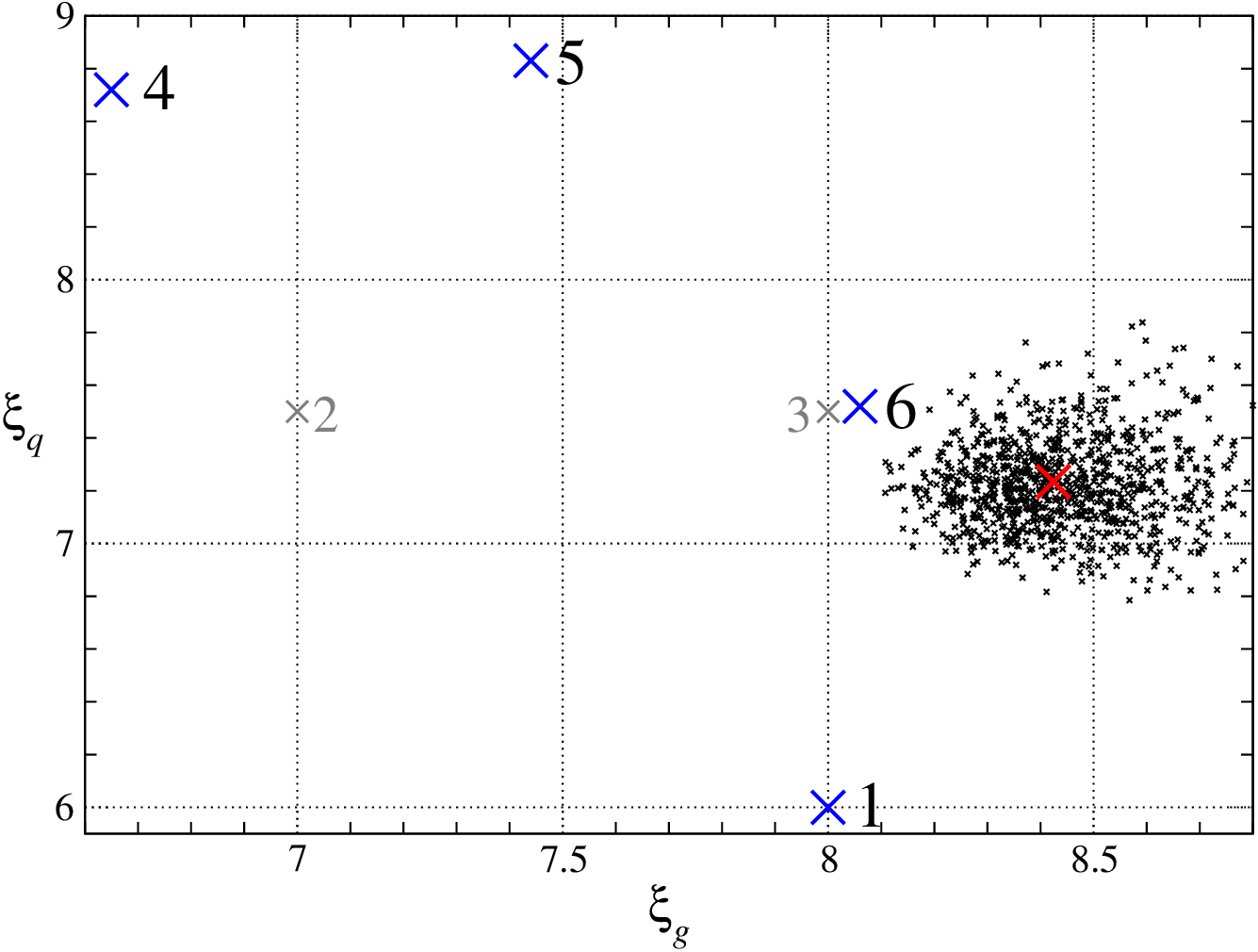}
\caption{As in Fig~\ref{fig:tuning-RT}. The figure shows the results from a plane fit 
using parameters from runs 1, 4, 5 and 6 (marked with an $\times$).
The big red (gray) cross at $(\xi^0_g,\xi^0_q)=(8.42,7.43)$ indicates
the result of the best fit.}
\label{fig:scatter}
\end{figure}

\begin{figure}
\includegraphics*[width=\colw]{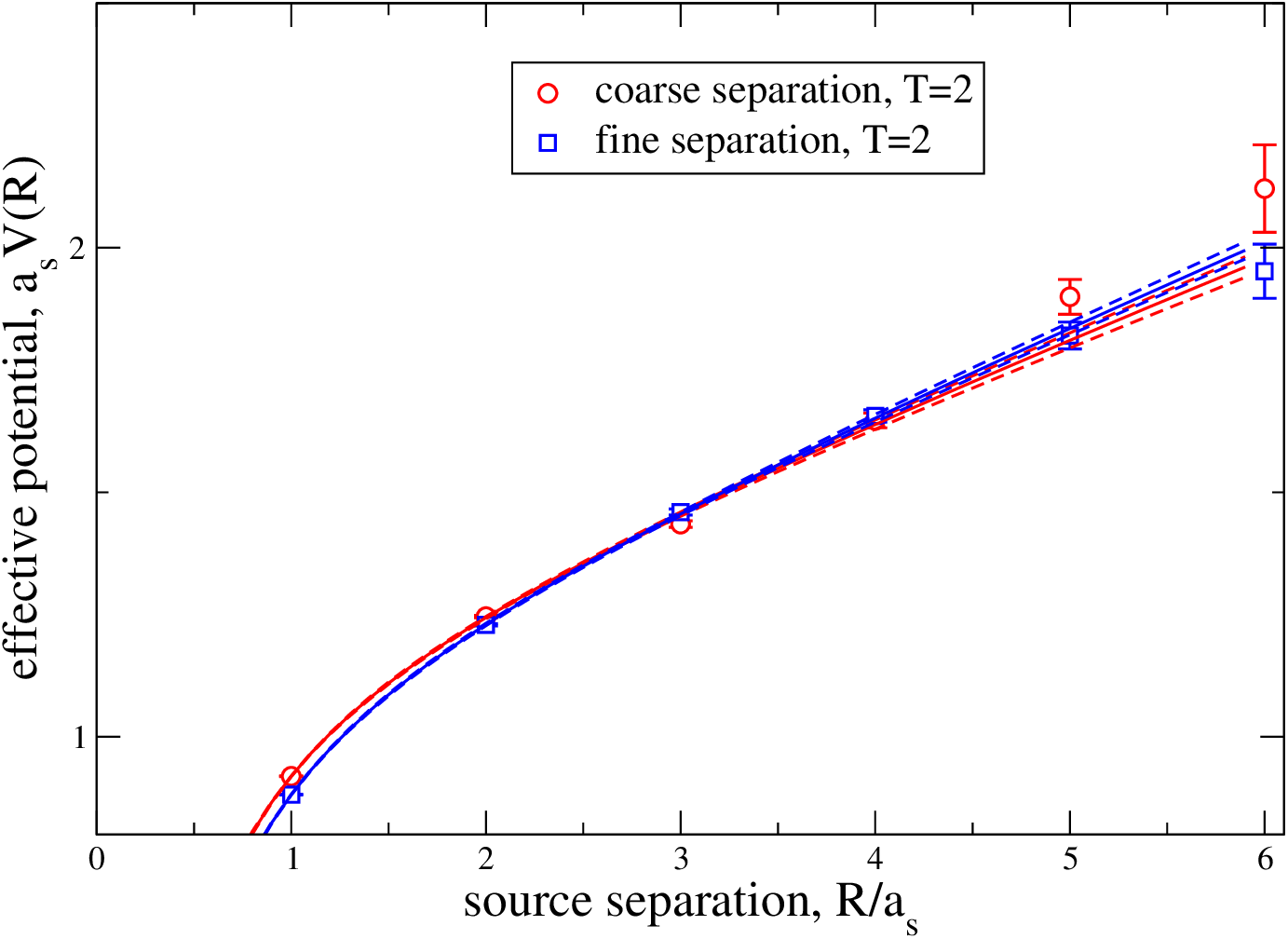}
\caption{The potential between fundamental static color sources for
  run 6, measured from
static propagation in a coarse direction. Lines show fits to the Cornell
potential, and are used in a crude determination of the lattice spacing.
\label{fig:potential}}
\end{figure}

We repeated the plane fit procedure including the new information 
from Run 6. Figure~\ref{fig:scatter} shows the resulting scatterplot determined 
on the $8^3\times80$ lattice from runs 1, 4, 5 and 6. The intersection points 
shift in a direction to move $c_g$ and $\xi_q$ even closer to the target anisotropy.
In order to get a rough idea of the physical scales of these lattices,
we compute the pion mass, the rho mass and the string tension.  We find
$a_tm_\pi=0.066(1)$ and $a_tm_\rho=0.120(5)$, which gives $m_\pi/m_\rho=0.54$, 
while a crude measurement (shown in Fig.~\ref{fig:potential}) of the 
string tension gives $a_s=0.18$fm.  
A more precise
determination of the lattice spacing will be obtained from the 1P--1S
splitting in charmonium~\cite{Juge:2005nr}.

\section{Conclusions}
\label{sec:conclusions}

We have performed a first simulation of 2-flavour QCD with improved 
Wilson fermions on anisotropic
lattices, with both quark and gluon anisotropies tuned to $\xi=6$
\footnote{While this paper was in preparation the results of a 
dynamical anisotropic simulation using staggered fermions appeared in
\cite{Levkova:2006gn}.}.
The tuning was based on a linear Ansatz for the dependence of
renormalised anisotropies on bare anisotropy parameters in a region of
parameter space.  The results from the final run demonstrate that the
tuning procedure, described in Sec.~\ref{sec:method}, works
satisfactorily.

The final, tuned point was found to lie marginally outside the
triangle used for the plane fit procedure, so the end result was based
on an extrapolation rather than an interpolation.  This increases both
the statistical and systematic uncertainties of the determination.  To
avoid this problem, it is important to choose a large enough triangle
to start with, so that successive parameter determinations are always
based on interpolations.

We also found that the original ($8^3\times48$) lattices used were too
short in the time direction to allow a reliable determination of
ground state energies, which were found to be systematically high, in
particular for higher momenta.  This led in turn to systematically
high values for $\xi_q$.  The adoption of lattices with longer time
extent was a crucial step in the procedure.  As
Table~\ref{tab:fitranges} shows, the optimal fit ranges were generally
found to be beyond the range of the shorter lattice.

We were able to determine the tuned parameters $(\xi^0_g,\xi^0_g)$
with a statistical uncertainty of 1\% and 3\% respectively from our ensembles of 250
configurations.  In addition, there are three main sources of
systematic uncertainties:
\begin{enumerate}
\item The $R$ and $T$ values used in the determination of the sideways
  potential, and the fit ranges used in the determination of the
  pseudoscalar dispersion relation.  Since the fit ranges are chosen to
  give stable ground state energies, we can safely assume that the
  latter is a small effect.  The effect of varying $R$ is also small, as
  shown in Fig.~\ref{fig:tuning-RT}. There may be a systematic error
  arising from the choice of $T$, but this
  is obscured by the larger statistical uncertainties in the $T=3$
  data, particularly in the $\xi^0_g$ direction. 
\item Lattice sizes.  The pion dispersion relation is unlikely to be
  strongly affected by the finite lattice volume, but the static quark
  potential may contain finite volume errors which affect our
  results.  We will be performing simulations at the tuned point on
  larger volumes, which will show whether this is a significant
  issue.
\item Nonlinearities in the dependence of $(\xi_g,\xi_q)$ on
  $(\xi^0_g,\xi^0_q)$.  Our final fit to four points shows no evidence
  of any significant nonlinearity.  If this were found to be a
  serious issue in any future simulation, a two-step procedure may be
  adopted where a smaller triangle centred on the preliminary tuned
  point is used in the second step.
\end{enumerate}
We have yet to verify that we get the same quark anisotropy from other
hadronic probes, for example the vector meson.  Differences in the
anisotropies can arise from lattice artefacts and can thus be
considered part of the finite lattice spacing errors.

These lattices will in the future be employed for a wide range of
physics investigations, including charm physics and heavy exotics
\cite{Juge:2005nr}, spectral functions at high temperature
\cite{Morrin:2005zq}, static--light mesons and baryons
\cite{Foley:2005af}, strong decays and flavour singlets including
glueballs.  These studies will be carried out on larger lattice
volumes.  Simulations on finer lattices will necessitate a new
nonperturbative tuning process like the one performed here; this will
be desirable in the longer term.

\begin{acknowledgments}
This work was supported by the IITAC project, funded by the Irish Higher Education Authority 
under PRTLI cycle 3 of the National Development Plan
and funded by 
IRCSET award SC/03/393Y, SFI grants 04/BRG/P0266 and 04/BRG/P0275.  
We are grateful to the Trinity Centre for High-Performance Computing for their
support and would like to thank Colin Morningstar for generous access to 
computing resources in the physics department of Carnegie Mellon University in
the early stages of this work.

\end{acknowledgments}

\bibliography{trinlat}

\begin{thebibliography}{10}

\bibitem{Karsch:1982ve}
F.~Karsch,
\newblock Nucl. Phys. {\bf B205}, 285 (1982).

\bibitem{Manke:1998qc}
CP-PACS, T.~Manke {\em et~al.},
\newblock Phys. Rev. Lett. {\bf 82}, 4396 (1999), [hep-lat/9812017].

\bibitem{Morningstar:1999rf}
C.~J. Morningstar and M.~J. Peardon,
\newblock Phys. Rev. {\bf D60}, 034509 (1999), [hep-lat/9901004].

\bibitem{Asakawa:2003re}
M.~Asakawa and T.~Hatsuda,
\newblock Phys. Rev. Lett. {\bf 92}, 012001 (2004), [hep-lat/0308034].

\bibitem{Hashimoto:2003fs}
S.~Hashimoto and M.~Okamoto,
\newblock Phys. Rev. {\bf D67}, 114503 (2003), [hep-lat/0302012].

\bibitem{Ishii:2005vc}
N.~Ishii, T.~Doi, Y.~Nemoto, M.~Oka and H.~Suganuma,
\newblock Phys. Rev. {\bf D72}, 074503 (2005), [hep-lat/0506022].

\bibitem{Foley:2004jf}
TrinLat, J.~Foley, A.~{\'O}~Cais, M.~Peardon and S.~M. Ryan,
\newblock Phys. Rev. {\bf D73}, 014514 (2006), [hep-lat/0405030].

\bibitem{Alford:2000an}
M.~G. Alford, I.~T. Drummond, R.~R. Horgan, H.~Shanahan and M.~J. Peardon,
\newblock Phys. Rev. {\bf D63}, 074501 (2001), [hep-lat/0003019].

\bibitem{Klassen:1998ua}
T.~R. Klassen,
\newblock Nucl. Phys. {\bf B533}, 557 (1998), [hep-lat/9803010].

\bibitem{Morningstar:1999dh}
C.~Morningstar and M.~J. Peardon,
\newblock Nucl. Phys. Proc. Suppl. {\bf 83}, 887 (2000), [hep-lat/9911003].

\bibitem{Morningstar:2003gk}
C.~Morningstar and M.~J. Peardon,
\newblock Phys. Rev. {\bf D69}, 054501 (2004), [hep-lat/0311018].

\bibitem{Morrin:2005tc}
R.~Morrin, M.~Peardon and S.~M. Ryan,
\newblock PoS {\bf LAT2005}, 236 (2005), [hep-lat/0510016].

\bibitem{Peardon:2002sd}
M.~J. Peardon,
\newblock Nucl. Phys. Proc. Suppl. {\bf 109A}, 212 (2002).

\bibitem{Duane:1987de}
S.~Duane, A.~D. Kennedy, B.~J. Pendleton and D.~Roweth,
\newblock Phys. Lett. {\bf B195}, 216 (1987).

\bibitem{Foley:2005ac}
J.~Foley {\em et~al.},
\newblock Comp. Phys. Commun. {\bf 172}, 145 (2005), [hep-lat/0505023].

\bibitem{Juge:2005nr}
K.~J. Juge, A.~{\'O}~Cais, M.~B. Oktay, M.~J. Peardon and S.~M. Ryan,
\newblock PoS {\bf LAT2005}, 029 (2005), [hep-lat/0510060].

\bibitem{Morrin:2005zq}
R.~Morrin {\em et~al.},
\newblock PoS {\bf LAT2005}, 176 (2005), [hep-lat/0509115].

\bibitem{Foley:2005af}
J.~Foley {\em et~al.},
\newblock PoS {\bf LAT2005}, 216 (2005), [hep-lat/0511005].

\bibitem{Levkova:2006gn}
L.~Levkova, T.~Manke and R.~Mawhinney,
\newblock Phys. Rev. {\bf D73}, 074504 (2006), [hep-lat/0603031].

\end{thebibliography}

\end{document}